\documentstyle[aps,twocolumn,prl,epsf]{revtex}

\begin{document}

\newcommand{\be}{\begin{equation}}
\newcommand{\ee}{\end{equation}}

\draft

\twocolumn[\hsize\textwidth\columnwidth\hsize\csname @twocolumnfalse\endcsname

\title{Theory of the ac Hall response of a model with X-ray edge Singularities
in infinite dimensions}
\author{Mukul S. Laad and Luis Craco}

\address{Max-Planck Institut fuer Physik komplexer Systeme, 38, Noethnitzer 
Stra${\beta}$e, 01187 Dresden, Germany}
\date{\today}
\maketitle

\widetext

\begin{abstract}
\noindent

We study the ac magnetotransport in a non-Fermi liquid metal, which possesses
explicit x-ray edge singularities in $d=\infty$.  Specifically, we compute the
ac conductivity tensor in a formalism that becomes exact in this limit.  The
ac Hall constant and Hall angle reveal features that are in striking
qualitative 
agreement with those observed in optical transmission experiments carried out
on $YBa_{2}Cu_{3}O_{7}$ thin films.  Our results provide a concrete realization
of the two-relaxation time picture proposed to explain magnetotransport
anomalies 
in the normal state of cuprate superconductors. 
\end{abstract}

]

\narrowtext
\section{INTRODUCTION}

  Discovery of cuprate superconductors has renewed interest in the physics
of strongly correlated electronic systems.  Extensive experimental work [1,2]
performed by several groups demonstrates that the normal state physical
 properties are not
understandable in terms of the Landau theory of the Fermi liquid (FL) [3],
which has 
been the mainstay of the conventional theory of metals.     

 In particular, dc transport in the {\it ab} plane in the normal state has 
been cited as evidence for the non-FL nature of these systems.  The inplane 
resistivity is linear in $T$ from temperatures as low as 10K to above room
temperature 
with no sign of saturation.  Hall measurements have provided more
striking evidence of the anomalous charge transport; the inverse Hall angle 
has a quadratic temperature dependence over a wide range in many materials [4].
The ac conductivity derived from optical reflectivity experiments shows a 
distinctly non-Drude like behavior above 200${\it cm^{-1}}$.  Thus the normal 
state transport reveals {\it two} qualitatively different relaxation rates  
associated with the decay of the usual transport and Hall currents in the
 system.  

  Quite a number of ideas have been proposed to explain these unusual
observations.  Very early on, almost seven years ago, Anderson proposed the
"two-relaxation-rates" scenario to understand these results.  In his theory
[5], the different relaxation rates arise as a consequence of spin-charge 
separation; the $T$-linear resistivity results from holons scattering off the
thermally excited spinons, whose number $\simeq T$, while spinon-spinon 
scattering in a magnetic field (which is like a fermion scattering process), 
gives rise to the quadratic Hall relaxation rate.  Carrington {\it et al.} have
proposed an alternative explanation in terms of differing relaxation rates 
and effective  masses associated with the anisotropic 2d Fermi surfaces 
in these materials [6].  However, as argued convincingly by Coleman [7], 
the FL-based theory suffers from a serious drawback; addition of impurities 
or underdoping changes the Fermi surface topology and 
results in a change in the $T$-dependence of $\theta_{H}^{-1}$.  This is not 
what is observed; these changes are observed to lead to a change in the $T$-
dependence of the resistivity, but the $T^{2}$-dependence of the Hall angle is
unchanged [4].  Theories based on the concept of singular skew-scattering in a
marginal Fermi liquid [8] share the above difficulty.

Drew {\it et al.} [9] have recently measured the far-infra-red (FIR)
transmission 
from thin films of $YBa_{2}Cu_{3}O_{7}$.  Their experimental results are 
consistent with a simple finite-frequency generalization of the Anderson-Ong
equation previously used to describe dc Hall effect [5].  However, the 
underlying hypothesis of the tomographic Luttinger liquid behavior in the 
metallic phase of the 2d Hubbard model is still open, inspite of intense 
importance of the issue.

  Theoretically, the problem of computing the transport coefficients for a 
strongly correlated fermionic system in a controlled way is a rather hard task.
The problem is even harder in the case of magnetotransport; to compute the Hall
conductivity tensor, one has to evaluate explicitly three-point functions.
In finite spatial dimensions, vertex corrections, which may be important, 
cannot be evaluated satisfactorily in any controlled approximation.  The above 
difficulties make it imperative to search for controlled approximation schemes
where some of the above difficulties can be circumvented without sacrificing 
essential correlation effects.       

  The dynamical mean-field approximation (DMFA or $d=\infty$) has proved to be
a successful tool to investigate transport in strongly correlated systems in a
controlled way [10].  This is because the vertex corrections entering in the 
Bethe-Salpeter eqn for the two-particle propagator for the conductivity vanish
rigorously in $d=\infty$.  To evaluate the conductivity tensor, one needs
only to compute the fully interacting {\it local} self-energy of the given model,
following which the Kubo formalism can be employed [11].  Given that DMFA 
captures the nontrivial local dynamics exactly, one expects that it provides 
an adequate physical description in situations where local fluctuations are
dominant.

  Lange [12] has employed the Mori-Zwanzig projection formalism to study dc
magnetotransport in the Hubbard model.  The actual evaluation of the
complicated equations is, however, actually carried out in the Hubbard I approximation.
This is known to lead to spurious instabilities (like 
ferromagnetism, which is washed away when local quantum fluctuations 
are included).  Moreover, only the dc Hall constant is evaluated explicitly.  
The Hubbard I approximation does not correctly capture the transfer of 
high-energy 
spectral weight to low energy upon hole doping, a feature characteristic of 
correlated systems, and so one expects that it will be inadequate when 
one attempts to look at the ac conductivity.  Majumdar {\it et al.} [13] have 
computed the Hall constant in the $d=\infty$ Hubbard model with next-nearest 
neighbor (nnn) hopping.  It is known, however, that the quantum paramagnetic
metallic phase of the Hubbard model is always a Fermi liquid  
in $d=\infty$ [10], and so their 
results should be applicable only at temperatures above $T_{coh}$, below which
local FL behavior sets in.  The computation of the ac magnetotransport for a 
non-FL in $d=\infty$ has not been attempted at all; only the ac conductivity
has been computed [14].  
           
 In this letter, we address precisely this issue.  We study the ac magnetotransport 
expected in the paramagnetic (non-FL) metallic state of a model which has
explicit x-ray edge singularities in $d=\infty$ [15].  Our motivation for the 
choice of this model is two-fold.  Firstly, it has been the first model to have
been solved exactly in $d=\infty$ [10], and has an explicit non-FL metal phase.
The second motivation comes from the proposal due to Anderson [3], who argues 
that the non-FL metallic behavior in the normal state of cuprates arises from 
processes which lead to  
 an {\it effective} blocking of the $\downarrow$-spin recoil in the 2d HM near 
half-filling.  The resulting x-ray edge (XRE) singularities drive the non-FL
features observed.  In view of the above discussion, we expect that an exact 
(in $d=\infty$) computation of the ac conductivity and Hall response in an
appropriate model (the imbedded impurity model with XRE singularities) should 
capture the essential features if the non-FL behavior is indeed driven by XRE
effects.  

\section{COMPUTATION OF THE AC HALL EFFECT}

We start with the simplified Hubbard model (SHM),
\be
H = -t\sum_{<ij>}(c_{i\uparrow}^{\dag}c_{j\uparrow}+h.c) + 
U\sum_{i}n_{i\uparrow}n_{i\downarrow} - \mu \sum_{i}(n_{i\uparrow}+n_{i\downarrow})
\ee
where the $\downarrow$-electrons do not hop.  A model of a similar type has been
 proposed by Sire {\it et al.} [16] as an {\it effective} model derived from a 
full three-band Hubbard model.  Interestingly enough, this is also (locally)
equivalent
to an effective model obtained from the HM if the $\downarrow$-spin recoil is
suppressed.  As is known, this model possesses an explicit analytical solution 
in $d=\infty$.  In our work below, we neglect the actual 2d character of the 
$Cu-O$ planes, and work with a semielliptic unperturbed DOS. 
An assumption implicit in the choice here is that bandstructure 
effects are not important, and that the origin of the non-FL behavior is
entirely due to strong correlation effects.  As observed by Shastry {\it et
al.} [4], the Hall constant in a strongly correlated system is dominated by
spectral weight far from the Fermi surface, and hence is independent of its
shape.  This lends good support to the assumption made above.
   To compute the ac conductivity 
tensor in $d=\infty$, one needs only the full s.p propagator 
$G_{\uparrow}({\bf k},\omega)=1/(\omega-\epsilon_{\bf k}-\Sigma_{\uparrow}(\omega))$, 
where $\Sigma_{\uparrow}(\omega)$ is the exact, local self-energy of the model eqn.(1) 
in $d=\infty$.  Since this is already known, we do not repeat the derivation
here, but refer the reader to [17].
Given $G_{\uparrow}({\bf k},\omega)$, one can compute the full interacting DOS.  This is 
the only input to the optical conductivity, calculated within the framework of 
the Kubo approach [10] in $d=\infty$.  The off-diagonal conductivity is, 
however, harder to compute.  
The Hamiltonian in a magnetic field is

\be
H=-\sum_{<ij>}t_{ij}({\bf A})(c_{i\uparrow}^{\dag}c_{j\uparrow}+h.c) + 
U\sum_{i}n_{i\uparrow}n_{i\downarrow}
\ee
with \mbox{$<ij>$} denoting neartest neighbors on a Bethe lattice in $d=\infty$.  The 
hopping matrix elements are modified by a Peierls phase factor and are
\mbox{$t_{ij}=exp(2i\pi/\phi_{0}\int_{i}^{j}{\bf A}\cdot d{\bf l})$}, where {\bf A} is 
the vector potential and $\phi_{0}=hc/e$.  The off-diagonal part of the 
conductivity involves the computation of a {\it three-point} function to first order in
the external field, as mentioned before [18]. Fortunately, a convenient form has been
worked out by Lange [12], so we use the approach developed there.  Explicitly,
after a somewhat tedious calculation, the imaginary part of
$\sigma_{xy}(\omega)$ is given by

\begin{eqnarray}
\nonumber
\sigma_{xy}^{"}(\omega) &= & c_{xy}\int_{-\infty}^{+\infty} d\epsilon
\rho_{0}(\epsilon)\epsilon \int_{-\infty}^{+\infty} d\omega_{1}
d\omega_{2} A_{\uparrow}(\epsilon,\omega_{1})A_{\uparrow}(\epsilon,\omega_{2})
\\ &&\frac{1}{\omega} \left[
\frac{F(\epsilon,\omega_{1};\omega)-
F(\epsilon,\omega_{2};\omega)}{\omega_{1}-\omega_{2}}
+(\omega \rightarrow -\omega) \right]
\end{eqnarray}
where

\be
F(\epsilon, \omega; \omega_{1}) = A_{\uparrow}(\epsilon, \omega_{1}-\omega)
[f(\omega_{1})-f(\omega_{1}-\omega)]
\ee
and $A_{\uparrow}(\epsilon,\omega)
=-Im [\omega-\epsilon-\Sigma_{\uparrow}(\omega)]^{-1}/\pi$ is the
s.p spectral function in $d=\infty$.  

\section{RESULTS AND DISCUSSION}

We now describe the results of our calculation.  The calculations were
performed with a semielliptic DOS with an effective half-width $D(=0.1 ev)$ 
and $U/D=20$ which 
\begin{figure}[ht]
\epsfxsize=3.5in
\epsffile{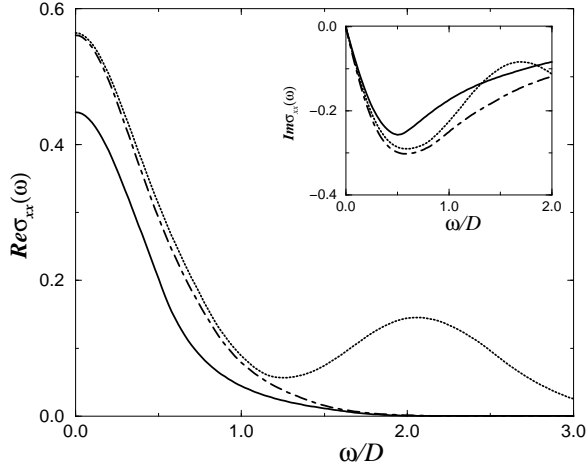}
\caption{Real and Imaginary (in inset) parts of the optical conductivity 
$\sigma_{xx}$ for $U/D=20$, $\delta=0.1$ (continuous) and $\delta=0.2$ 
(dot-dashed). The dotted line shows the case with $U/D=2$, $\delta=0.2$.  
The low-energy behavior is practically unchanged in the two cases.
Notice the non-Drude 
fall-off at intermediate frequency ($\omega \approx 0.2eV$). Inset 
shows the imaginary part of $\sigma_{xx} (\omega) \approx 
\omega^\gamma$, $\gamma=0.9$.}
\label{fig1}
\end{figure}
\hspace*{-.3cm}reproduces an insulating solution at half-filling.
We specify that this effective bandwidth is not obviously related to the "free"
bandwidth taken from spectroscopic estimates, but should be thought of as an 
independent parameter of our proposed model.  We have checked (see fig.1) 
that decreasing $U$ ($U/D=2$) does not affect the low energy part of the
$\sigma_{xx}(\omega)$, which continues to have the same frequency dependence
upto $\omega/D \simeq 1$.  
\begin{figure}[bh]
\epsfxsize=3.5in
\epsffile{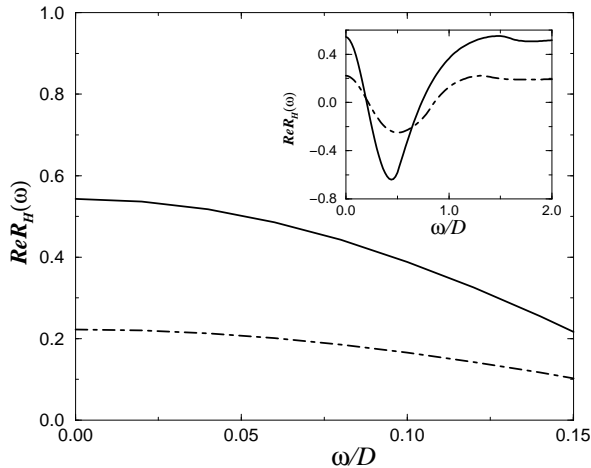}
\caption{$Re R_H (\omega)$ for  $U/D=20$,
$\delta=0.1$ (continuous) and $\delta=0.2$ 
(dot-dashed). The full frequency dependence 
is shown in the inset. $Re R_H(\omega)$ falls off much faster with 
$\omega$ in comparison to $Re \sigma_{xx} (\omega)$
and is qualitatively in agreement with the experimental 
results of Ref. [9]. } 
\label{fig2}
\end{figure}
\begin{figure}[htb]
\epsfxsize=3.5in
\epsffile{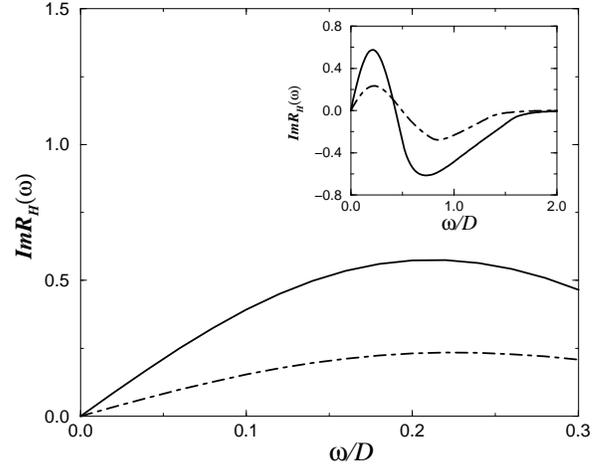}
\caption{$Im R_H (\omega)$ for  $U/D=20$,
$\delta=0.1$ (continuous) and $\delta=0.2$ 
(dot-dashed). The full frequency dependence 
is shown in the inset. Notice the qualitative agreement with 
the experimental results of Ref.~[9].} 
\label{fig3}
\end{figure}
\hspace*{-.3cm}All calculations were performed at a temperature $T=0.01D$.  
The computed optical conductivity $\sigma_{xx}(\omega)$ has a distinctly 
non-Drude fall off at intermediate 
frequencies, while Im$\sigma_{xx}(\omega) \simeq \omega^{\gamma}$ 
at small $\omega$ with $\gamma =0.9$ for both 
$n=1-\delta = 0.9, 0.8$ with our parameters (Fig.~\ref{fig1}).  More
interesting, however, is the result of the computation of the ac Hall effect.
In Fig.~\ref{fig2}, we show the real part of the ac Hall constant as
a function of frequency.   The corresponding imaginary part is depicted  in
Fig.~\ref{fig3}. The results show striking similarities to the thin-film
data of Drew {\it et al.}.  It is interesting to notice that the frequency 
 dependence of the $R_{H}^{'}(\omega)$ is much faster 
$(\simeq \omega^{2\gamma})$  than that of
$\sigma_{xx}(\omega)$, and thus our results show that in this frequency 
regime, the ac transport is indeed characterized
by two relaxation rates describing relaxation of longitudinal and Hall currents
in the system.  Additional support is provided by results for the 
ac Hall angle. We show the real part of cot$\theta_{H}(\omega)$ in
Fig.~\ref{fig4}; the frequency dependence of the real and the 
imaginary parts of cot$\theta_{H}(\omega)$ observed experimentally is qualitatively 
reproduced by our calculation.  Finally, the real part of
tan$\theta_{H}(\omega)$ is depicted in Fig.~\ref{fig5}.
Clearly, Re$tan\theta_{H}(\omega)$ falls off
much faster ($\simeq \omega^{2\gamma}$)
with $\omega$ than does $\sigma_{xx}(\omega)$, providing
additional evidence [7] in favor of the two-relaxation time scenario.
Calculation of the dc Hall effect [19] using the FKM shows that near $n=1$,
the resistivity goes linearly with $T$, while cot$\theta_{H}(T)\simeq aT^{2}+b$
as one would expect if the "two-relaxation time" scenario were valid.
Additionally, these distinct behaviors persist as a function of hole 
doping; the curves are qualitatively similar
 for $\delta=0.1, 0.2$.  Thus, our calculation shows
unambiguously that an exact (in $d=\infty$) treatment of the local dynamical
fluctuations in a model with infrared singularities qualitatively reproduces
the essential features of ac magnetotransport observed in 
\begin{figure}[t]
\epsfxsize=3.5in
\epsffile{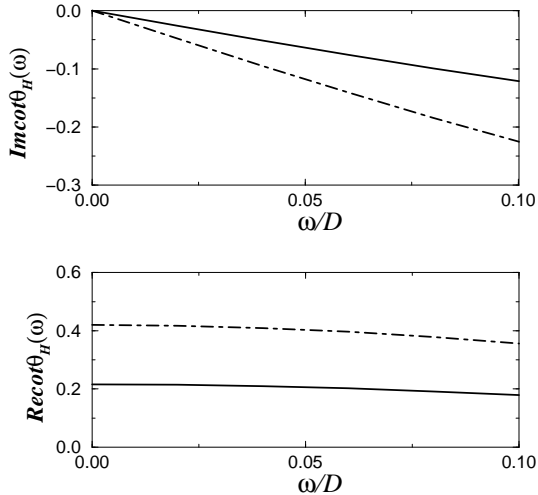}
\caption{Real and Imaginary parts of $cot \theta_H (\omega)$
for  $U/D=20$, $\delta=0.1$ (continuous) and $\delta=0.2$ 
(dot-dashed). The agreement with experimental results of
Ref. [9] is striking.} 
\label{fig4}
\end{figure}
\hspace*{-.3cm}experiment; the
longitudinal and transverse responses to an applied Lorentz field are governed
by qualitatively different timescales.

It is instructive to comment here that our results are very different from
those expected for a Fermi liquid.  In the $d=\infty$ Hubbard model, which is
a Fermi liquid below a certain crossover scale $T_{coh}$, the low-$T$
magnetotransport would necessarily be characterized by a {\it single} relaxation
rate expected from FL transport [6].  Furthermore, the imaginary part of the
optical conductivity computed above (fig.1, inset) shows appreciable frequency
dependence, contrary to expectations from a FL picture where it is expected to
be $\omega$-independent.  In ref.[14], it has been found that away from the
half-filled case, Im$\Sigma_{\uparrow}(\omega) \simeq -const + a|\omega-\mu|$
at low energy.  We have also computed the frequency dependence of the transverse
 relaxation rate in a magnetic field using
\be
\Gamma_{xy}(\omega)=\frac{\omega\sigma_{xy}'
(\omega)}{\sigma_{xy}''(\omega)} \;.
\ee
We find that $\Gamma_{xy}(\omega)$ goes quadratically with $\omega$ at 
low energy.  The resistivity is determined by the longitudinal scattering rate, i.e 
by Im$\Sigma_{\uparrow}(\omega)$, and so goes like $\rho(T) \simeq aT+b$, while 
the quadratic $T$-dependence of $\Gamma_{xy}(\omega)$ implies that $cot \theta_{H}(T)
\simeq c_{1}T^{2}+c_{2}$, as found earlier [19]. 
 Thus, our results represent a concrete realization of the idea of the
``two-relaxation rates'' picture proposed to understand magnetotransport 
anomalies in the ``normal'' state of cuprate superconductors.
 The non-Fermi liquid character of the charge transport is also shown up in the
frequency dependent Hall effect.  In a Fermi liquid, $R_{H}'(\omega)$ is
expected to be a constant, independent of $\omega,T$, while $R_{H}''(\omega)=0$.
Clearly, this is very different from our results summarized in figs(2-5).

\begin{figure}[t]
\epsfxsize=3.5in
\epsffile{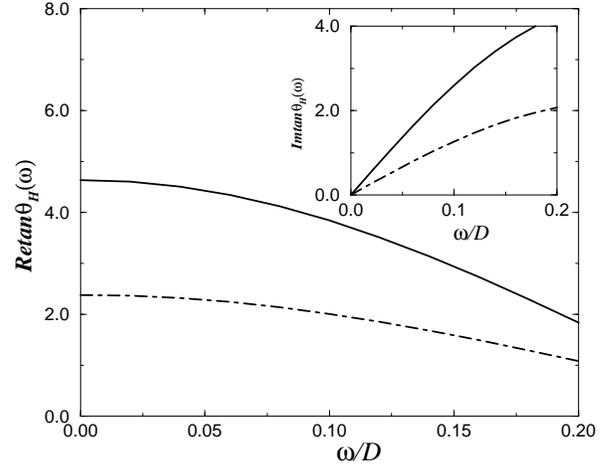}
\caption{Real and Imaginary (in inset) 
parts of $tan \theta_H (\omega)$
for  $U/D=20$, $\delta=0.1$ (continuous) and $\delta=0.2$ 
(dot-dashed). The real part shows qualitative agreement 
with results from Ref. [9]. } 
\label{fig5}
\end{figure}

  In view of the qualitative agreement of our obtained results with
the experimental results of Drew {\it et al.}, we conclude that the charge
dynamics is that consistent with a picture where FL theory is invalidated near
half-filling in a model with low-energy XRE singularities.  In this model, the
low-energy response is scale-invariant, in contrast to what is found in the
$d=\infty$ Hubbard model (a FL for $T < T_{coh}$); in fact the only scale is the
 temperature, as in the marginal FL theory.

 It is interesting to inquire about a possible link with Anderson's ideas.
To see this, we notice that the impurity version of the model eqn.(1) is 
precisely the x-ray edge model.  This was bosonized many years ago 
by Schotte {\it et al.} [20], and the resulting low energy behavior was found 
to be the same as that of a Tomonaga-Luttinger model defined on a half-line.  
Upon refermionization, the resulting model is effectively one-dimensional 
in each radial direction around the ``impurity'' (recall that in $d=\infty$, the
lattice model is mapped onto a selfconsistently embedded single impurity 
model).  This is very similar to Anderson's tomographic Luttinger liquid 
ideas, and the link is suggestive.  We have solved the selfconsistently 
embedded (lattice) version of the XRE model to study the effects of low-energy
singularities on the charge dynamics. Our results indicate that the anomalous 
features observed in ac magnetotransport are understandable in terms of 
the dynamics of the non-FL metallic state in a model characterized by XRE 
singularities.

To conclude, we have presented, we believe, the first controlled calculation
of the ac magnetotransport for a non-FL metal in $d=\infty$.  Our calculated 
results reveal that the longitudinal and transverse responses in a non-FL are 
governed by distinct relaxation time-scales, in agreement with the
two-relaxation time 
picture proposed to explain the dc magnetotransport in cuprates.  Qualitative
agreement with published far-infrared data on YBCO thin films is seen, and so our
results are also an evidence for the importance of X-ray edge-like effects in
the ``normal'' non-FL metallic state of the doped cuprate superconductors. 

\acknowledgments

One of us (MSL) thanks the Alexander von Humboldt 
Stiftung for financial support.  We thank Prof. P. Fulde for advice and 
hospitality at the MPIPKS, Dresden.

\end{document}